\documentclass[10pt]{article}
\usepackage{amssymb,amsmath}
\usepackage{amsfonts}
\usepackage{eufrak}
\usepackage{graphicx}

\begin{document}
\begin{center}
{\large {\bf Transverse Wave Propagation in Relativistic Two-fluid\\
Plasmas around Schwarzschild-de Sitter Black Hole}}\\
\vspace{2cm}

M. Hossain Ali \footnote {\it The Abdus Salam International Centre for Theoretical Physics, Strada Costiera 11, 34014 Trieste, Italy.
E-mail: $m_-hossain_-ali_-bd@yahoo.com$ (Corresponding author).} and
M. Atiqur Rahman\footnote{\it E-mail: atirubd@yahoo.com}\\
{\it Department of Applied Mathematics, University of Rajshahi, Rajshahi-6205, Bangladesh.}
\end{center}
\vspace{3cm} \centerline{\bf Abstract} \baselineskip=18pt
\bigskip
We investigate transverse electromagnetic waves propagating in a plasma influenced by the gravitational field of the Schwarzschild-de Sitter black hole. Applying 3+1 spacetime split we derive the relativistic two-fluid equations to take account of gravitational effects due to the event horizon and describe the set of simultaneous linear equations for the perturbations. We use a local approximation to investigate the one-dimensional radial propagation of Alfv\'en and high frequency electromagnetic waves. We derive the dispersion relation for these waves and solve it for the wave number $k$ numerically.
\vspace{0.5cm}\\
{\it PACS number(s)}: 95.30.Sf, 95.30.Qd, 97.60.Lf

\vfill

\newpage

\section{Introduction}\label{sec1}
Black holes are still mysterious \cite{one}. Physicists are grappling the theory of black holes, while astronomers are searching for real-life examples of black holes in the universe \cite{two}. There still exists no convincing observational data which can prove conclusively the existence of black holes in the universe. Black holes, however, are not objects of direct observing. They must be observed indirectly through the effects they exert on their environment. It is proved that black holes exist on the basis of study of effects which they exert on their surroundings.

Black holes greatly affect the surrounding plasma medium (which is highly magnetized) with their enormous gravitational fields. Hence plasma physics in the vicinity of a black hole has become a subject of great interest in astrophysics. In the immediate neighborhood of a black hole general relativity applies. It is therefore of interest to formulate plasma physics problems in the context of general relativity.

A covariant formulation of the theory based on the fluid equations of general relativity and Maxwell's equations in curved spacetime has so far proved unproductive because of the curvature of four-dimensional spacetime in the region surrounding a black hole. Thorne, Price, and Macdonald (TPM) \cite{three,four,five,six} developed a method of a 3 + 1 formulation of general relativity in which the electromagnetic equations and the plasma physics at least look somewhat similar to the usual formulations in flat spacetime while taking accurate account of general relativistic effects such as curvature. In the TPM formulation, work connected with black holes has been facilitated by replacing the hole's event horizon with a membrane endowed with electric charge, electrical conductivity, and finite temperature and entropy. The membrane paradigm is mathematically equivalent to the standard, full general relativistic theory of black holes so far as physics outside the event horizon is concerned. But the formulation of all physics in this region turns out to be very much simpler than it would be using the standard covariant approach of general relativity.

Buzzi, et.al. \cite{seven,eight}, using the TPM formalism described a general relativistic version of two-fluid formulation of plasma physics and investigated the nature of plasma waves (transverse waves in \cite{seven}, and longitudinal waves in \cite{eight}) near the horizon of the Schwarzschild black hole. In this paper we apply the formalism of Buzzi et.al. \cite{seven} to investigate the transverse electromagnetic waves propagating in a plasma close to the Schwarzschild black hole in the de Sitter (dS) space, called the Schwarzschild-de Sitter (SdS) black hole. The study is interesting because of the presence of cosmological constant ($\Lambda$). There has been a renewed interest in cosmological constant as it is found to be present in the inflationary scenario of the early universe. In these scenario the universe undergoes a stage where it is geometrically similar to dS space \cite{nine}. Among other things inflation has led to the cold dark matter. According to the cold dark matter theory, the bulk of the dark matter is in the form of slowly moving particles (axions or neutralinos). If the cold dark matter theory proves correct, it would shed light on the unification of forces \cite{ten,eleven}. Comprehensive review works of the cosmological constant or dark energy, including the observational evidence of it and the problems associated with it, have been done by many authors \cite{twelve,thirteen,fourteen,fifteen,sixteen,seventeen,eighteen,nineteen,twenty}.

In recent years, considerable attention has been concentrated on the study of black holes in dS spaces. This is motivated, basically, by the following two aspects: first, several different types of astrophysical observations indicate that our universe is in a phase of accelerating expansion \cite{twenty one,twenty two,twenty three,twenty four,twenty five,twenty six,twenty seven,twenty eight,twenty nine,thirty,thirty one,thirty two,thirty three,thirty four,thirty five,thirty six,thirty seven,thirty eight,thirty nine,fourty,fourty one,fourty two,fourty three,fourty four,fourty five,fourty six}. Associated with this acceleration is a positive cosmological constant. Our universe therefore might approach a dS phase in the far future. Second, similar as the AdS/CFT correspondence, an interesting proposal, the so-called dS/CFT correspondence, has been suggested that there is a dual relation between quantum gravity on a dS space and Euclidean conformal field theory (CFT) on a boundary of dS space \cite{fourty seven,fourty eight,fourty nine,fifty,fifty one,fifty two}. In view of the above reasons, our study of transverse wave propagation in relativistic two-fluid plasma in the environment close to the event horizon of the SdS black hole is interesting. The result we have obtained reduces to that of the Schwarzschild black hole \cite{seven} when the cosmological constant vanishes.

This paper is organized as follows. In section \ref{sec2}, we summarize the 3+1 formulation of general relativity. In section \ref{sec3}, we describe the nonlinear two-fluid equations expressing continuity and conservation of energy and momentum. The two-fluids are coupled together through Maxwell's equations for the electromagnetic fields. In section \ref{sec4}, we restrict one-dimensional wave propagation in the radial $z$ (Rindler coordinate system) direction, and linearize the equations for wave propagation in section \ref{sec5} by giving a small perturbation to fields and fluid parameters. We express the derivatives of the unperturbed quantities with respect to $z$. In section \ref{sec6}, we discuss the local or mean-field approximation used to obtain numerical solutions for the wave dispersion relations. We describe the dispersion relation for the transverse waves. We give the numerical procedure for determining the roots of the dispersion relation in section \ref{sec7}. In section \ref{sec8} we present the numerical solutions for the wave number $k$. Finally, in section \ref{sec9}, we present our remarks. We use units $G=c=k_B=1$.

\section{Formalism in 3+1 Spacetime}\label{sec2}
In this section we apply TPM formulation to split the Schwarzschild-de Sitter black hole spacetime, which is the solution of Einstein equations with a positive $\Lambda(=3/\ell^2)$ term corresponding to a vacuum state spherical symmetric configuration. The metric of the spacetime is asymptotically de Sitter and has the form
\begin{eqnarray}
ds^2&=&g_{\mu \nu }dx^\mu dx^\nu\nonumber\\
 &=&-\Delta^2dt^2  +\frac{1}{\Delta^2}dr^2+r^2(d\theta^2+\textrm{sin}^2\theta d\varphi^2),\label{eq1}
\end{eqnarray}
where the metric function is
\begin{equation}
 \Delta^2=1-\frac{2M}{r}-\frac{r^2}{\ell^2}, \label{eq2}
\end{equation}
$M$ being the mass of the black hole and the coordinates are defined such that $-\infty\leq t\leq \infty $, $r\geq 0$, $0\leq \theta \leq
\pi $, and $0\leq \phi \leq 2\pi$. At large $r$, the metric (\ref{eq1}) tends to the dS space limit. The explicit dS case is obtained by setting $M=0$ while the explicit Schwarzchild case is obtained by taking the limit $\ell\rightarrow\infty$. When $\ell^2$ is replace by $-\ell^2$, the metric (\ref{eq1}) describes an interesting nonrotating AdS black hole called the Schwarzschild Anti-de Sitter (SAdS) black hole. The AdS black holes play a major role in the AdS/CFT correspondence \cite{fifty three,fifty four,fifty five,fifty six} and they have also received interest in the context of brane-world scenarios based on the setup of Randall and Sundrum \cite{fifty seven,fifty eight}.

The horizons of the SdS black hole are located at the real positive roots of $\Delta^2(r)\equiv\frac{1}{\ell^2r}(r-r_h)(r-r_c)(r_--r)=0$, and there are more than one horizon if $0<\Xi<1/27$ where $\Xi=M^2/\ell^2$. The black hole (event) horizon $r_h$ and the cosmological horizon $r_c$ are located, respectively, at \begin{eqnarray}
r_h&=&\frac{2M}{\sqrt{3\Xi}}\textrm{cos}\frac{\pi+\psi}{3}\label{eq3},\\
r_c&=&\frac{2M}{\sqrt{3\Xi}}\textrm{cos}\frac{\pi-\psi}{3}\label{eq4},
\end{eqnarray}
where
\begin{equation}
\psi=\textrm{cos}^{-1}(3\sqrt{3\Xi})\label{eq5}.
\end{equation}
The negative root
\begin{equation}
r_-=-\frac{2M}{\sqrt{3\Xi}}\textrm{cos}\frac{\psi}{3}\label{eq6}
\end{equation}
has no physical meaning. In the limit $\Xi\rightarrow0$, one finds that $r_h\rightarrow 2M$ and $r_c\rightarrow \ell$, and it is obvious that $r_c>r_h$, i.e., the event horizon is the smallest positive root. The spacetime is dynamic for $r<r_h$ and $r>r_c$. The two horizons coincide: $r_h=r_c=3M$ (extremal), when $\Xi=1/27$, and the spacetime then becomes the well known Nariai spacetime. Expanding $r_h$ in terms of $M$ with $\Xi<1/27$, we obtain
\begin{equation}
r_h\approx2M\left(1+\frac{4M^2}{\ell^2}+\cdot \cdot\cdot\right)\label{eq7},
\end{equation}
that is, the event horizon of the SdS black hole is greater than the Schwarzschild event horizon, $r_{Sch}=2M$. For $\Xi>1/27$, the spacetime is dynamic for all $r>0$, that is, the metric (\ref{eq1}) then represents not a black hole but an unphysical naked singularity at $r=0$.

The hypersurfaces of constant universal time $t$ define an absolute three-dimensional space described by the metric
\begin{equation}
ds^2=g_{ij}dx^idx^j=\frac{1}{\Delta^2}dr^2+r^2(d\theta ^2+{\rm
sin}^2\theta d\varphi ^2)\label{eq8},
\end{equation}
where the indices $i, j$ range over $1,2,3$ and refer to coordinates in absolute space. The fiducial observers (FIDOs), the observers at rest with respect to this absolute space, measure their proper time $\tau $ using clocks that they carry with them and make local measurements of physical quantities. Then all their measured quantities are defined as FIDO locally measured quantities and all rates measured by them are measured using FIDO proper time. The FIDOs use a local Cartesian coordinate system with unit basis vectors tangent to the coordinate lines
\begin{equation}
{\textbf{e}}_{\textbf{r}}=\Delta\frac{\partial }{\partial r},\qquad{\textbf{e}}_{\hat \theta }=\frac{1}{r}\frac{\partial }{\partial \theta },\qquad{\textbf{e}}_{\hat \varphi }=\frac{1}{r{\rm sin}\theta }\frac{\partial }{\partial \varphi }\label{eq9}.
\end{equation}
For a spacetime viewpoint this set of orthonormal vectors also includes the basis vector for the time coordinate, given by \begin{equation}
{\textbf{e}}_{\hat 0}=\frac{d}{d\tau }=\frac{1}{\alpha }\frac{\partial }{\partial t}\label{eq10},
\end{equation}
where $\alpha $ is the lapse function (or redshift factor) defined by
\begin{equation}
\alpha (r)\equiv \frac{d\tau }{dt}=\left(1-\frac{2M}{r}-\frac{r^2}{\ell^2}\right)^{\frac{1}{2}}\label{eq11}.
\end{equation}
The FIDO proper time $\tau$ functions as a local laboratory time, where the FIDOs have the role of \lq \lq local laboratories\rq \rq . The coordinate time $t$ slices spacetime in the way that the FIDOs would do physically.

The lapse function $\alpha$, which governs the ticking rates of clocks and redshifts, plays the role of a gravitational potential. It also provides the gravitational acceleration felt by a FIDO \cite{three,four,five,six}:
\begin{equation}
{\textbf{a}}=-\nabla {\rm ln}\alpha =-\frac{1}{\alpha }\left({\frac{M}{r^2}}-{\frac{r}{\ell^2}}\right){\textbf{e}}_{\hat r}\label{eq12}.
\end{equation}
The rate of change of any scalar physical quantity or any three-dimensional vector or tensor, as measured by a FIDO, is defined by the derivative
\begin{equation}
\frac{D}{D\tau }\equiv \left(\frac{1}{\alpha }\frac{\partial }{\partial t}+{\bf v}\cdot \nabla \right)\label{eq13},
\end{equation}
$\textbf{v}$ being the fluid velocity measured locally by a FIDO.

\section{Two-Fluid Equations in 3+1 Formalism}\label{sec3}
We consider two-component plasma such as an electron-positron plasma or electron-ion plasma. The continuity equation for each of the fluid species in the 3+1 notation is
\begin{equation}
\frac{\partial }{\partial t}(\gamma _sn_s)+\nabla \cdot (\alpha \gamma _sn_s{\textbf{v}}_s)=0\label{eq14},
\end{equation}
where $s$ is 1 for electrons and 2 for positrons (or ions). For a perfect relativistic fluid of species $s$ in three-dimensions, the energy density $\epsilon _s$, the momentum density ${\textbf{S}}_s$, and stress-energy tensor $W_s^{jk}$ are given by
\begin{equation}
\epsilon _s=\gamma _s^2(\varepsilon _s+P_s{\textbf{v}}_s^2),\qquad{\textbf{S}}_s=\gamma _s^2(\varepsilon _s+P_s){\textbf{v}}_s,\qquad W_s^{jk}=\gamma _s^2(\varepsilon
_s+P_s)v_s^jv_s^k+P_sg^{jk}\label{eq15},
\end{equation}
where $\textbf{v}_s$ is the fluid velocity, $n_s$ is the number density, $P_s$ is the pressure, and $\varepsilon _s$ is the total energy density defined by
\begin{equation}
\varepsilon _s=m_sn_s+P_s/(\gamma _g-1)\label{eq16},
\end{equation}
the gas constant $\gamma _g$ being $4/3$ as $T\rightarrow \infty $ and $5/3$ as $T\rightarrow 0$.

The ion temperature profile is closely adiabatic and it approaches $10^{12}\,K$ near the horizon \cite{fifty nine}. Far from the event horizon, electron (positron) temperatures are essentially equal to the ion temperatures. But the electrons closer to the horizon are progressively cooled to about $10^8-10^9\,K$ by mechanisms such as multiple Compton scattering and synchrotron radiation. Using the conservation of entropy one can express the equation of state in the form:
\begin{equation}
\frac{D}{D\tau }\left(\frac{P_s}{n_s^{\gamma _g}}\right)=0\label{eq17},
\end{equation}
where $D/D\tau =(1/\alpha )\partial /\partial t+{\textbf{v}}_s\cdot \nabla $. For a relativistic fluid, the full equation of state as measured in the fluid's rest frame, is given by \cite{sixty,sixty one}:
\begin{equation}
\varepsilon =m_sn_s+m_sn_s\left[\frac{P_s}{m_sn_s}-\frac{{\rm i}H_2^{(1)^\prime }({\rm i}m_sn_s/P_s)}{H_2^{(1)}({\rm i}m_sn_s/P_s)}\right]\label{eq18},
\end{equation}
where the $H_2^{(1)}(x)$ are Hankel functions.

The fluid quantities in (\ref{eq15}) take the following form in the electromagnetic field:
\begin{eqnarray}
\epsilon _s&=&\frac{1}{8\pi }({\bf E}^2+{\textbf{B}}^2),\qquad{\textbf{S}}_s=\frac{1}{4\pi }{\textbf{E}}\times {\bf B}\nonumber,\\
W_s^{jk}&=&\frac{1}{8\pi }({\textbf{E}}^2+{\textbf{B}}^2)g^{jk}-\frac{1}{4\pi }(E^jE^k+B^jB^k)\label{eq19}.
\end{eqnarray}
Energy and momentum conservation equations are given by \cite{three,four,five}
\begin{eqnarray}
\frac{1}{\alpha }\frac{\partial }{\partial t}\epsilon _s&=&-\nabla \cdot {\textbf{S}}_s+2{\textbf{a}}\cdot {\textbf{S}}_s,\label{eq20}\\
\frac{1}{\alpha }\frac{\partial }{\partial t}{\textbf{S}}_s&=&\epsilon _s{\textbf{a}}-\frac{1}{\alpha }\nabla \cdot (\alpha {\stackrel{\leftrightarrow }{\textbf{W}}}_s)\label{eq21}.
\end{eqnarray}
When the two-fluid plasma couples to the electromagnetic fields, Maxwell's equations take the following 3+1 form:
\begin{eqnarray}
\nabla \cdot {\textbf{B}}&=&0,\label{eq22}\\
\nabla \cdot {\textbf{E}}&=&4\pi \sigma ,\label{eq23}\\
\frac{\partial {\textbf{B}}}{\partial t}&=&-\nabla \times (\alpha {\textbf{E}}),\label{eq24}\\
\frac{\partial {\textbf{E}}}{\partial t}&=&\nabla \times (\alpha {\textbf{B}})-4\pi \alpha {\textbf{J}},\label{eq25}
\end{eqnarray}
where the charge and current densities are respectively defined by
\begin{equation}
\sigma =\sum_s\gamma _sq_sn_s,\hspace{1.2cm}{\bf J}=\sum_s\gamma _sq_sn_s{\bf v}_s\label{eq26}.
\end{equation}

Using (\ref{eq16}) and (\ref{eq22}--\ref{eq25}), the energy and momentum conservation equations (\ref{eq20}) and (\ref{eq21}) can be rewritten for each species $s$ in the form
\begin{equation}
\frac{1}{\alpha }\frac{\partial }{\partial t}P_s-\frac{1}{\alpha }\frac{\partial }{\partial t}[\gamma _s^2(\varepsilon _s+P_s)]-\nabla \cdot [\gamma _s^2(\varepsilon _s+P_s){\textbf{v}}_s] +\gamma _sq_sn_s{\bf E}\cdot {\textbf{v}}_s+2\gamma _s^2(\varepsilon _s+P_s){\textbf{a}}\cdot {\textbf{v}}_s=0,\label{eq27}
\end{equation}
and
\begin{eqnarray}
\gamma _s^2(\varepsilon _s+P_s)\left(\frac{1}{\alpha }\frac{\partial }{\partial t}+{\textbf{v}}_s\cdot \nabla \right){\textbf{v}}_s+\nabla P_s -\gamma _sq_sn_s({\textbf{E}}+{\textbf{v}}_s\times {\textbf{B}})\nonumber\\
+{\textbf{v}}_s\left(\gamma _sq_sn_s{\textbf{E}}\cdot {\textbf{v}}_s+\frac{1}{\alpha }\frac{\partial }{\partial t}P_s\right)+\gamma _s^2(\varepsilon _s+P_s)[{\textbf{v}}_s({\textbf{v}}_s\cdot {\textbf{a}})-{\textbf{a}}]=0\label{eq28},
\end{eqnarray}
respectively. Although these equations are valid in a FIDO frame, they reduce for $\alpha=1$ to the corresponding special relativistic equations \cite{sixty two} which are valid in a frame in which both fluids are at rest. The transformation from the FIDO frame to the comoving (fluid) frame involves a boost velocity, which is simply the freefall velocity onto the black hole, given by
\begin{equation}
v_{\textrm{ff}}=(1-\alpha ^2)^{\frac{1}{2}}\label{eq29}.
\end{equation}
The relativistic Lorentz factor then becomes $\gamma
_{\textrm{boost}}\equiv (1-v_{\textrm{ff}}^2)^{-1/2}=1/\alpha $.

The two-fluid equations in SdS coordinates form the basis of the numerical procedure for solving the linear two-fluid equations; but they cannot be evaluated analytically. However, the Rindler coordinate system, in which space is locally Cartesian, provides a good approximation to the SdS metric near the event horizon. Without the complication of explicitly curved spatial three-geometries, the essential features of the horizon and the 3+1 split are preserved.

The SdS metric (\ref{eq1}) in Rindler coordinates is approximated by
\begin{equation}
ds^2=-\alpha^2dt^2+dx^2+dy^2+dz^2,\label{eq30}
\end{equation}
where
\begin{equation}
x=r_h\left(\theta -\frac{\pi }{2}\right),\qquad y=r_h\varphi ,\qquad z=2r_h\Delta\label{eq31}.
\end{equation}
The standard lapse function in Rindler coordinates becomes $\alpha =z/2r_h$, where $r_h$ is the event horizon of the black hole.

\section{Radial Wave Propagation in One-Dimension}\label{sec4}

We consider one-dimensional wave propagation in the radial $z$ direction. Introducing the complex variables
\begin{eqnarray}
v_{sz}(z,t)=u_s(z,t),\quad v_s(z,t)=v_{sx}(z,t)+\textrm{i}v_{sy}(z,t),\nonumber\\
B(z,t)=B_x(z,t)+\textrm{i}B_y(z,t), \quad E(z,t)=E_x(z,t)+\textrm{i}E_y(z,t)\label{eq32},
\end{eqnarray}
and setting
\begin{eqnarray}
v_{sx}B_y-v_{sy}B_x=\frac{\textrm{i}}{2}(v_sB^\ast -v_s^\ast B),\nonumber\\
v_{sx}E_y-v_{sy}E_x=\frac{\textrm{i}}{2}(v_sE^\ast -v_s^\ast E)\label{eq33},
\end{eqnarray}
where the $\ast $ denotes the complex conjugate, one could write the continuity equation (\ref{eq14}) in the form
\begin{equation}
\frac{\partial }{\partial t}(\gamma _sn_s)+\frac{\partial }{\partial z}(\alpha \gamma _sn_su_s)=0,\label{eq34}
\end{equation}
and Poisson's equation (\ref{eq23}) as
\begin{equation}
\frac{\partial E_z}{\partial z}=4\pi (q_1n_1\gamma _1+q_2n_2\gamma _2).\label{eq35}
\end{equation}

From the $\textbf{e}_{\hat{x}}$ and $\textbf{e}_{\hat{y}}$ components of (\ref{eq24}) and (\ref{eq25}), one could obtain
\begin{eqnarray}
\frac{1}{\alpha }\frac{\partial B}{\partial t}&=&-{\textrm{i}} \left(\frac{\partial }{\partial z}-a\right)E,\label{eq36}\\
{\textrm{i}}\frac{\partial E}{\partial t}&=&-\alpha \left(\frac{\partial }{\partial z}-a\right)B-{\textrm{i}}4\pi e\alpha (\gamma _2n_2v_2-\gamma _1n_1v_1)\label{eq37}.
\end{eqnarray}
Differentiating (\ref{eq37}) with respect to $t$ and using (\ref{eq36}), we obtain
\begin{equation}
\left(\alpha ^2\frac{\partial ^2}{\partial z^2}+\frac{3\alpha }{2r_h}\frac{\partial }{\partial z}-\frac{\partial ^2}{\partial t^2}+\frac{1}{(2r_h)^2}\right)E =4\pi e\alpha \frac{\partial }{\partial t}(n_2\gamma _2v_2-n_1\gamma _1v_1)\label{eq38}.
\end{equation}
The transverse component of the momentum conservation equation is obtained from the $\textbf{e}_{\hat{x}}$ and $\textbf{e}_{\hat{y}}$ components of (\ref{eq28}) as follows:
\begin{equation}
\rho _s\frac{Dv_s}{D\tau }=q_sn_s\gamma _s(E-\textrm{i}v_sB_z+\textrm{i}u_sB)-u_sv_s\rho _sa-v_s\left(q_sn_s\gamma _s\textbf{E}\cdot \textbf{v}_s+\frac{1}{\alpha }\frac{\partial P_s}{\partial t}\right),\label{eq39}
\end{equation}
where
\[
\textbf{E}\cdot \textbf{v}_s=\frac{1}{2}(Ev_s^\ast +E^\ast v_s)+E_zu_s,
\]
and $\rho _s$ is the total energy density defined by
\begin{equation}
\rho _s=\gamma _s^2(\varepsilon _s+P_s)=\gamma _s^2(m_sn_s+\Gamma _gP_s)\label{eq40}
\end{equation}
with $\Gamma _g=\gamma _g/(\gamma _g-1)$.

\section{Linearization}\label{sec5}

We apply perturbation method to linearize the equations derived in the preceding section. We introduce the quantities
\begin{eqnarray}
u_s(z,t)&=&u_{os}(z)+\delta u_s(z,t),\qquad v_s(z,t)=\delta
v_s(z,t),\nonumber\\
n_s(z,t)&=&n_{os}(z)+\delta n_s(z,t),\qquad P_s(z,t)=P_{os}(z)+\delta P_s(z,t),\nonumber\\
\rho _s(z,t)&=&\rho _{os}(z)+\delta \rho _s(z,t),\qquad \textbf{E}(z,t)=\delta \textbf{E}(z,t),\nonumber\\
\textbf{B}_z(z,t)&=&\textbf{B}_o(z)+\delta \textbf{B}_z(z,t),\qquad \textbf{B}(z,t)=\delta \textbf{B}(z,t)\label{eq41},
\end{eqnarray}
where an applied magnetic field has been chosen to lie along the radial $\textbf{e}_{\hat{z}}$ direction. The relativistic Lorentz factor is also linearized such that
\begin{equation}
\gamma _s=\gamma _{os}+\delta \gamma _s,\quad \mbox{where}\quad \gamma _{os}=\left(1-{\bf u}_{os}^2\right)^{-\frac{1}{2}},\quad \delta \gamma _s=\gamma _{os}^3{\bf u}_{os}\cdot \delta {\bf u}_s\label{eq42}.
\end{equation}
Near the event horizon the unperturbed radial velocity for each species as measured by a FIDO along $\textbf{e}_{\hat{z}}$ is assumed to be the freefall velocity so that
\begin{equation}
u_{os}(z)=v_{\textrm{ff}}(z)=[1-\alpha ^2(z)]^{\frac{1}{2}}\label{eq43}.
\end{equation}
It follows, from the continuity equation (\ref{eq34}), that
\[
r^2\alpha \gamma _{os}n_{os}u_{os}=\mbox{const.}=r_h^2\alpha _h\gamma _hn_hu_h,
\]
where the values with a subscript $h$ are the limiting values at the event horizon. The freefall velocity at the event horizon becomes unity so that $u_h=1$. Since $u_{os}=v_{\textrm{ff}}$, $\gamma _{os}=1/\alpha $; and hence $\alpha \gamma _{os}=\alpha _h\gamma _h=1$. Also, because $v_{\textrm{ff}}=(r_h/r)^{1/2}\chi$, the number density for each species can be written as follows:
\begin{equation}
n_{os}(z)=\frac{1}{\chi^4}n_{hs}v_{\textrm{ff}}^3(z)\label{eq44}, \end{equation}
where
\begin{equation}
\chi=\left(1-\frac{r^3}{r_hr_cr_-}\right)^{\frac{1}{2}} \left[1+r_h\left(\frac{1}{r_c}+\frac{1}{r_-} \right)\right]^{-\frac{1}{2}}\label{eq45}.
\end{equation}
When $\ell^2\rightarrow \infty$, (\ref{eq44}) reduce to the Schwarzschild result \cite{seven}. The equation of state (\ref{eq17}) and (\ref{eq44}) lead to write the unperturbed pressure in terms of the freefall velocity as
\begin{equation}
P_{os}(z)=\frac{1}{\chi^{4\gamma_g}}P_{hs}v_{\textrm{ff}}^{3\gamma _g}(z)\label{eq46}.
\end{equation}
Since $P_{os}=k_Bn_{os}T_{os}$, then with $k_B=1$ the temperature profile is
\begin{equation}
T_{os}(z)=\frac{1}{\chi^{4(\gamma_g-1)}}T_{hs}v_{\textrm{ff}}^{3(\gamma _g-1)}(z)\label{eq47}.
\end{equation}

The unperturbed magnetic field, chosen to be in the radial $z$ direction, is parallel to the infall fluid velocity $u_{os}(z)\textbf{e}_{\hat{z}}$ for each fluid. It thus does not experience effects of spatial curvature along with the infall fluid velocity. The magnetic field depends on the radial coordinate $r$ because of flux conservation. Since $\nabla \cdot \textbf{B}_o=0$, it follows that $r^2B_o(r)=\mbox{const.}$, from which one could obtain the unperturbed magnetic field in terms of the freefall velocity as
\begin{equation}
B_o(z)=\frac{1}{\chi^4}B_hv_{\textrm{ff}}^4(z)\label{eq48}.
\end{equation}
Since
\begin{equation}
\frac{dv_{\textrm{ff}}}{dz}=-\frac{\alpha }{2r_h}\frac{1}{v_{\textrm{ff}}},\label{eq49}
\end{equation}
we have
\begin{eqnarray}
\frac{du_{os}}{dz}&=&-\frac{\alpha }{2r_h}\frac{1}{v_{\textrm{ff}}},\qquad \frac{dB_o}{dz}=-\frac{4\alpha }{2r_h}\frac{B_o}{v_{\textrm{ff}}^2},\nonumber\\
\frac{dn_{os}}{dz}&=&-\frac{3\alpha }{2r_h}\frac{n_{os}}{v_{\textrm{ff}}^2},\qquad \frac{dP_{os}}{dz}=-\frac{3\alpha }{2r_h}\frac{\gamma _gP_{os}}{v_{\textrm{ff}}^2}\label{eq50}.
\end{eqnarray}

When the linearized variables from (\ref{eq41}) and (\ref{eq42}) are substituted into the continuity equation and products of perturbation terms are neglected, it follows that
\begin{eqnarray}
\gamma _{os}\left(\frac{\partial }{\partial t}+u_{os}\alpha \frac{\partial }{\partial z}+\frac{u_{os}}{2r_h}+\gamma _{os}^2\alpha \frac{du_{os}}{dz}\right)\delta n_s +\left(\alpha \frac{\partial }{\partial z}+\frac{1}{2r_h}\right)(n_{os}\gamma _{os}u_{os})\nonumber\\
+n_{os}\gamma _{os}^3\left[u_{os}\frac{\partial }{\partial t}+\alpha \frac{\partial }{\partial z}+\frac{1}{2r_h}+\alpha \left(\frac{1}{n_{os}}\frac{dn_{os}}{dz}+3\gamma _{os}^2u_{os}\frac{du_{os}}{dz}\right)\right]\delta u_s=0\label{eq51}.
\end{eqnarray}
The conservation of entropy, (\ref{eq17}), on linearization, gives
\begin{equation}
\delta P_s=\frac{\gamma _gP_{os}}{n_{os}}\delta n_s\label{eq52}.
\end{equation}
Then from the total energy density, (\ref{eq40}), it follows that
\begin{equation}
\delta \rho _s=\frac{\rho _{os}}{n_{os}}\left(1+\frac{\gamma _{os}^2\gamma _gP_{os}}{\rho _{os}}\right)\delta n_s+2u_{os}\gamma _{os}^2\rho _{os}\delta u_s,\label{eq53}
\end{equation}
where $\rho _{os}=\gamma _{os}^2(m_sn_{os}+\Gamma _gP_{os})$. Linearizing the transverse part of the momentum conservation equation, differentiating it with respect to $t$, and then substituting from (\ref{eq36}), we derive
\begin{equation}
\left(\alpha u_{os}\frac{\partial }{\partial z}+\frac{\partial }{\partial t}-\frac{u_{os}}{2r_h}+\frac{{\rm i}\alpha q_s\gamma _{os}n_{os}B_o}{\rho _{os}}\right)\frac{\partial \delta v_s}{\partial t}-\frac{\alpha q_s\gamma _{os}n_{os}}{\rho _{os}}\left(\alpha u_{os}\frac{\partial }{\partial z}+\frac{\partial }{\partial t}+\frac{u_{os}}{2r_h}\right)\delta E=0\label{eq54}.
\end{equation}
When linearized, Poisson's equation (\ref{eq35}) and (\ref{eq38}), respectively, give
\begin{equation}
\frac{\partial \delta E_z}{\partial z}=4\pi e(n_{o2}\gamma _{o2}-n_{o1}\gamma _{o1})+4\pi e(\gamma _{o2}\delta n_2-\gamma _{o1}\delta n_1)+4\pi e(n_{o2}u_{o2}\gamma _{o2}^3\delta u_2-n_{o1}u_{o1}\gamma _{o1}^3\delta u_1),\label{eq55}
\end{equation}
\begin{equation}
\left(\alpha ^2\frac{\partial ^2}{\partial z^2}+\frac{3\alpha }{2r_h}\frac{\partial }{\partial z}-\frac{\partial ^2}{\partial t^2}+\frac{1}{(2r_h)^2}\right)\delta E=4\pi e\alpha \left(n_{o2}\gamma _{o2}\frac{\partial \delta v_2}{\partial t}-n_{o1}\gamma _{o1}\frac{\partial \delta v_1}{\partial t}\right)\label{eq56}.
\end{equation}

\section{Dispersion Relation}\label{sec6}

We restrict our consideration to effects on a local scale for which the distance from the horizon does not vary significantly. We apply a local (or mean-field) approximation for the lapse function $\alpha(z)$ and hence for the equilibrium fields and fluid quantities. If the plasma is situated relatively close to the event horizon, $\alpha ^2\ll 1$, then a relatively small change in distance $z$ will make a significant difference to the magnitude of $\alpha $. Thus it is important to choose a sufficiently small range in $z$ so that $\alpha(z)$ does not vary much.

We consider thin layers in the $\textbf{e}_{\hat{z}}$ direction, each layer with its own $\alpha _o$, where $\alpha _o$ is some mean value of $\alpha $ within a particular layer. By considering a large number of layers within a chosen range of $\alpha _o$ values, a more complete picture can then be built up. However, such a local approximation imposes a restriction on the magnitude of the wavelength and, hence, on the wave number $k$. It is assumed that the wavelength is small in comparison with the range over which the equilibrium quantities change significantly. Then the wavelength must be smaller in magnitude than the scale of the gradient of the lapse function $\alpha $, i.e.,
\[
\lambda <\left(\frac{\partial \alpha }{\partial z}\right)^{-1}=2r_h\simeq \zeta 5.896\times 10^5\textrm{cm},
\]
or, equivalently,
\[
k>\frac{2\pi }{2r_h}\simeq \zeta^{-1}1.067\times
10^{-5}\textrm{cm}^{-1},\quad 1\leq\zeta\leq1.5,
\]
for a black hole of mass $\sim 1M_\odot $. The value for $\zeta=1.5$ corresponds to the extremal SdS black hole, while the value for $\zeta=1$ corresponds to the Schwarzschild black hole.

The hydrodynamical approach used in this work has some disadvantages, such as it is essentially a bulk, fluid approach. So the microscopic behavior of the two-fluid plasma is treated in a somewhat approximate manner via the equation of state. It means that the results are really only strictly valid in the long wavelength limit. However, the restriction, imposed by the local approximation, on the wavelength is not too severe and permits the consideration of intermediate to long wavelengths so that the small $k$ limit is still valid.

The unperturbed field and fluid quantities are assumed not to be constant with respect to $\alpha(z)$. Then, using the local approximation for $\alpha$, the derivatives of the equilibrium quantities can be evaluated at each layer for a given $\alpha_o$. In the local approximation for $\alpha $, $\alpha \simeq \alpha _o$ is valid within a particular layer. Hence, the unperturbed fields and fluid quantities and their derivatives, which are functions of $\alpha $, take on their corresponding \lq \lq mean-field\rq \rq values for a given $\alpha _0$. Then the coefficients in (\ref{eq51}), (\ref{eq54}), and (\ref{eq55}) become constants within each layer, evaluated at each fixed mean-field value, $\alpha=\alpha_o$. So it is possible to Fourier transform the equations with respect to $z$, using plane-wave-type solutions for the perturbations of the form $\sim e^{\textrm{i}(kz-\omega t)}$ for each $\alpha _o$ layer.

When Fourier transformed, (\ref{eq54}) and (\ref{eq56}) turn out to be
\begin{equation}
\omega \left(\alpha _oku_{os}-\omega +\frac{\textrm{i}u_{os}}{2r_h}+\frac{\alpha _oq_s\gamma _{os}n_{os}B_o}{\rho _{os}}\right)\delta v_s -\textrm{i}\alpha _o\frac{q_s\gamma _{os}n_{os}}{\rho _{os}}\left(\alpha _oku_{os}-\omega -\frac{\textrm{i}u_{os}}{2r_h}\right)\delta E=0 \label{eq57},
\end{equation}
\begin{equation}
\delta E=\frac{\textrm{i}4\pi e\alpha _o\omega (n_{o2}\gamma _{o2}\delta v_2-n_{o1}\gamma _{o1}\delta v_1)}{\alpha _ok(\alpha _ok-{\textrm{i}}3/2r_h)-\omega ^2-1/(2r_h)^2}\label{eq58}.
\end{equation}
The dispersion relation for the transverse electromagnetic wave modes may be put in the form
\begin{equation}
\left[K_\pm \left(K_\pm \pm \frac{\textrm{i}}{2r_h}\right)-\omega ^2+\frac{1}{(2r_h)^2}\right] =\alpha _o^2\left\{\frac{\omega _{p1}^2(\omega -u_{o1}K_\pm )}{(u_{o1}K_\mp -\omega -\alpha _o\omega _{c1})} +\frac{\omega _{p2}^2(\omega -u_{o2}K_\pm )}{(u_{o2}K_\mp -\omega +\alpha _o\omega _{c2})}\right\}\label{eq59}
\end{equation}
for either the electron-positron or electron-ion plasma, where $K_\pm =\alpha _ok\pm \textrm{i}/2r_h$, $\omega _{cs}=e\gamma _{os}n_{os}B_o/\rho _{os}$ and $\omega_{ps}=\sqrt{4\pi e^2\gamma^2_{os}n^2_{os}/\rho_{os}}$. Like the plasma frequency $\omega_{ps}$, the cyclotron frequency $\omega _{cs}$ is frame independent. Although the fluid quantities are measured in the fluid frame, the field $B_o$ is measured in the FIDO frame. So, the factors of $\gamma _{os}$ do not cancel out explicitly. The transformation $B_o\rightarrow \gamma _{os}B_o$ boosts the fluid frame for either fluid and thereby cancels the $\gamma _{os}$ factors. The \lq \lq $+$\rq \rq  and \lq \lq $-$\rq \rq  denote the left $L$ and right $R$ modes, respectively. The dispersion relation for the $L$ mode is obtained by taking the complex conjugate of the dispersion relation for the $R$ mode. Both of the two modes have the same dispersion relation in the special relativistic case.

\section{Numerical Solution of Modes}\label{sec7}
We have to solve the dispersion relation (\ref{eq59}) in order to determine all the physically meaningful modes for the transverse waves. But the dispersion relation is complicated enough, even in the simplest cases for the electron-positron plasma where both species are assumed to have the same equilibrium parameters, and an analytical solution is cumbersome and unprofitable. We therefore solve the dispersion relation numerically and for this purpose we put it in the form of a matrix equation as follows:
\begin{equation}
(A-kI)X=0\label{eq60},
\end{equation}
where the eigenvalue is chosen to be the wave number $k$, the eigenvector $X$ is given by the relevant set of perturbations, and $I$ is the identity matrix.

To write the perturbation equations in an appropriate form, we introduce the following set of dimensionless variables:
\begin{eqnarray}
&&\tilde \omega =\frac{\omega }{\alpha _o\omega _\ast },\quad \tilde k=\frac{kc}{\omega _\ast },\quad k_h=\frac{1}{2r_h\omega _\ast },\nonumber\\
&&\delta \tilde u_s=\frac{\delta u_s}{u_{os}},\quad \tilde v_s=\frac{\delta v_s}{u_{os}},\quad \delta \tilde n_s=\frac{\delta n_s}{n_{os}},\nonumber\\
&&\delta \tilde B=\frac{\delta B}{B_o},\quad \tilde E=\frac{\delta E}{B_o},\quad \delta \tilde E_z=\frac{\delta E_z}{B_o}.\label{eq61}
\end{eqnarray}
For an electron-positron plasma, $\omega _{p1}=\omega _{p2}$ and $\omega _{c1}=\omega _{c2}$, since the choice of input parameters is the same for each fluid. Hence, $\omega _\ast $ is defined as
\begin{equation}
\omega _\ast =\left\{\begin{array}{rl}&\omega _c \quad\mbox{Alfv\'en modes},\\
&\\
&(2\omega _p^2+\omega _c^2)^{\frac{1}{2}}\quad \mbox{high frequency modes},\end{array}\right.\label{eq62}
\end{equation}
where $\omega _p=\sqrt{\omega _{p1}\omega _{p2}}$ and $\omega _c=\sqrt{\omega _{c1}\omega _{c2}}$. However, for the case of an electron-ion plasma, the plasma frequency and the cyclotron frequency are different for each fluid, and so the choice of $\omega _\ast $ is a more complicated matter. For simplicity, it is assumed that
\begin{equation}
\omega _\ast =\left\{\begin{array}{rl}&\frac{1}{\sqrt{2}}(\omega _{c1}^2+\omega _{c2}^2)^{\frac{1}{2}}\quad \mbox{Alfv\'en modes},\\
&\\
&(\omega _{\ast 1}^2+\omega _{\ast 2}^2)^{\frac{1}{2}}\quad \mbox{high frequency modes},\end{array}\right.\label{eq63}
\end{equation}
where $\omega _{\ast s}^2=(2\omega _{ps}^2+\omega _{cs}^2)$. These chosen values for $\omega_\ast$ reduce to the special relativistic cutoffs in the zero gravity limit for an electron-positron plasma. The cutoffs, in the special relativistic case, are determined by the dispersion relation in view of the fact that the solutions to the dispersion relation are physical (i.e., Re$(k)>0$) only for certain frequency regimes. Since the dispersion relations cannot be handled analytically, it is difficult to determine the cutoffs in the case including gravity. Other similar combinations for $\omega_\ast$ make only a difference of a scale factor to the form of the results as $\omega_\ast$ is really.

We write the dimensionless eigenvector for the transverse set of equations as
\begin{equation}
\tilde X_{\rm transverse}=\left[\begin{array}{c}\delta \tilde v_1\\\delta \tilde v_2\\\delta \tilde B\\\delta \tilde E\end{array}\right]\label{eq64}.
\end{equation}
When linearized and Fourier transformed, equations (\ref{eq36}) and (\ref{eq37}) turn out to be
\begin{equation}
\left(k-\frac{\textrm{i}}{2r_h\alpha _o}\right)\delta E+\frac{\textrm{i}\omega }{\alpha _o}\delta B=0,\label{eq65}
\end{equation}
and
\begin{equation}
\frac{\textrm{i}\omega }{\alpha _o}\delta E=\left(k-\frac{\textrm{i}}{2r_h\alpha _o}\right)\delta B+4\pi e(\gamma _{o2}n_{o2}\delta v_2-\gamma _{o1}n_{o1}\delta v_1)\label{eq66}.
\end{equation}

Using (\ref{eq61}), we write (\ref{eq57}), (\ref{eq65}), and (\ref{eq66}) in the dimensionless form as follows:
\begin{eqnarray}
\tilde k\delta \tilde v_s&=&\left(\frac{\tilde \omega }{u_{os}}-\left(\frac{q_s}{e}\right)\frac{\omega _{cs}}{u_{os}\omega _\ast }-\frac{\textrm{i}k_h}{\alpha _o}\right)\delta \tilde v_s +\left(\frac{q_s}{e}\right)\frac{\omega _{cs}}{u_{os}\omega _\ast }\delta \tilde B-\textrm{i}\left(\frac{q_s}{e}\right)\frac{\omega _{cs}}{u_{os}\omega _\ast }\delta \tilde E,\label{eq67}\\
\tilde k\delta \tilde E&=&-\textrm{i}\tilde \omega \delta \tilde B+\frac{{\rm i}k_h}{\alpha _o}\delta \tilde E,\label{eq68}\\
\tilde k\delta \tilde B&=&u_{o1}\frac{\omega _{p1}^2}{\omega _{c1}\omega _\ast }\delta \tilde v_1-u_{o2}\frac{\omega _{p2}^2}{\omega _{c2}\omega _\ast }\delta \tilde v_2+\frac{\textrm{i}k_h}{\alpha _o}\delta \tilde B+\textrm{i}\tilde \omega \delta \tilde E\label{eq69}.
\end{eqnarray}
These are the equations in the required form to be used as input to (\ref{eq60}).

\section{Results}\label{sec8}
We have carried out the numerical analysis by using the well known MATLAB. We have considered both the electron-positron plasma and the electron-ion plasma. The limiting horizon values for the magnetic field and the fluid parameters are chosen as follows. For the electron-positron plasma, the horizon values are taken to be
\begin{equation}
n_{hs}=10^{18}{\textrm{cm}}^{-3},\quad T_{hs}=10^{10}\textrm{K},\quad B_h=3\times 10^6\textrm{G},\quad\mbox{and}\quad\gamma _g=\frac{4}{3}\label{eq70}.
\end{equation}

\begin{figure}[h]\label{fig1}
\begin{center}
\includegraphics[scale=0.33]{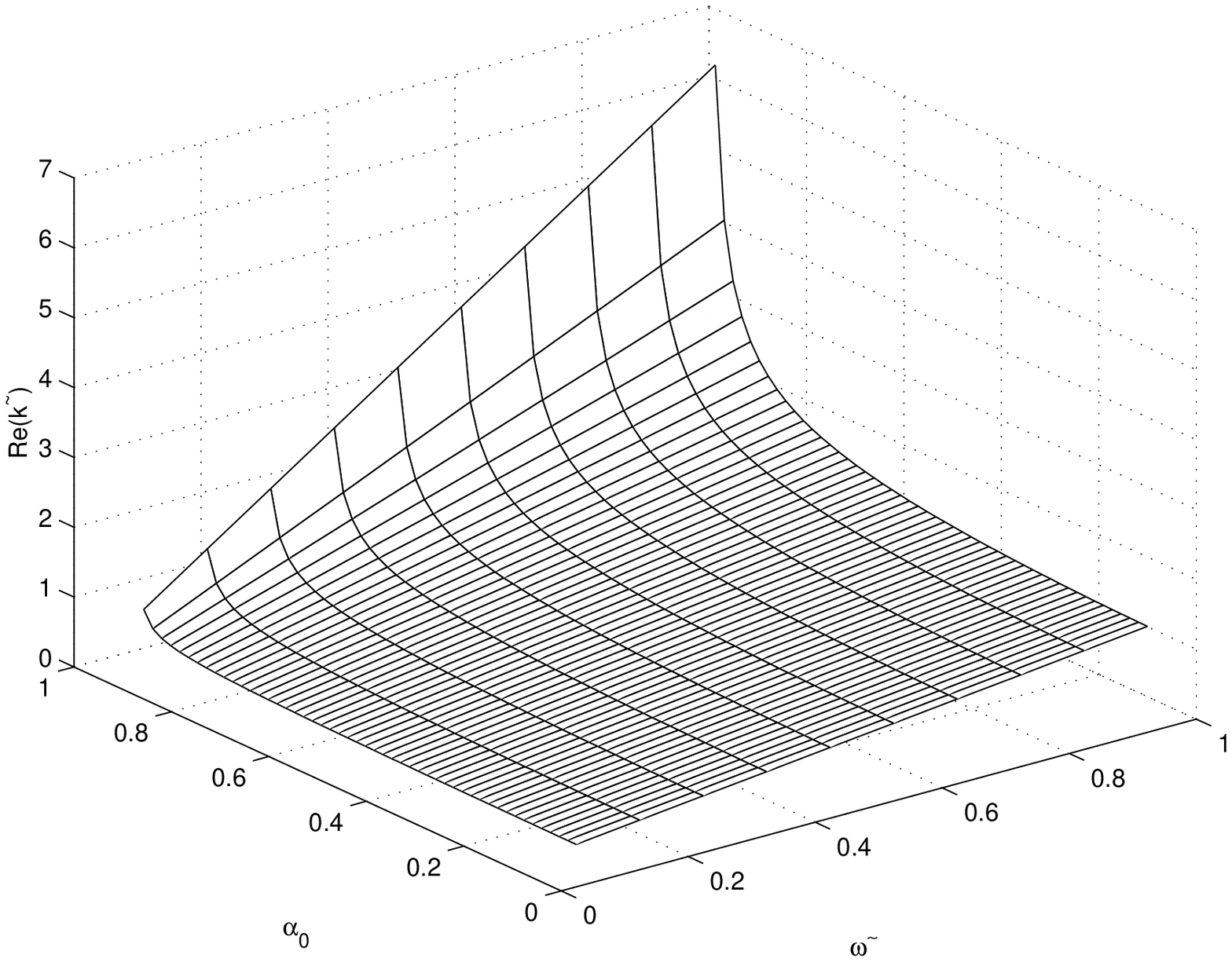}\\
\includegraphics[scale=0.33]{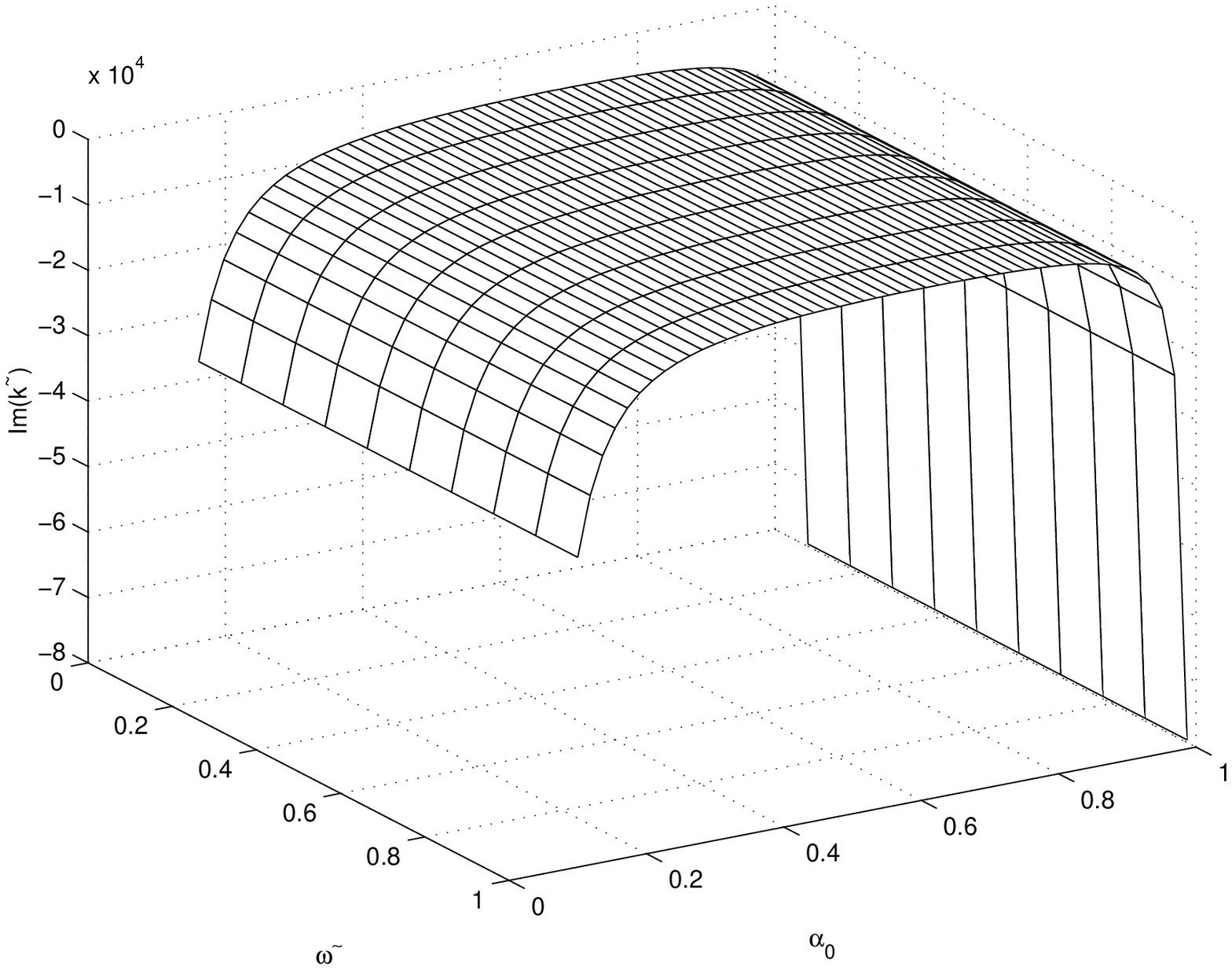}\\
\includegraphics[scale=0.33]{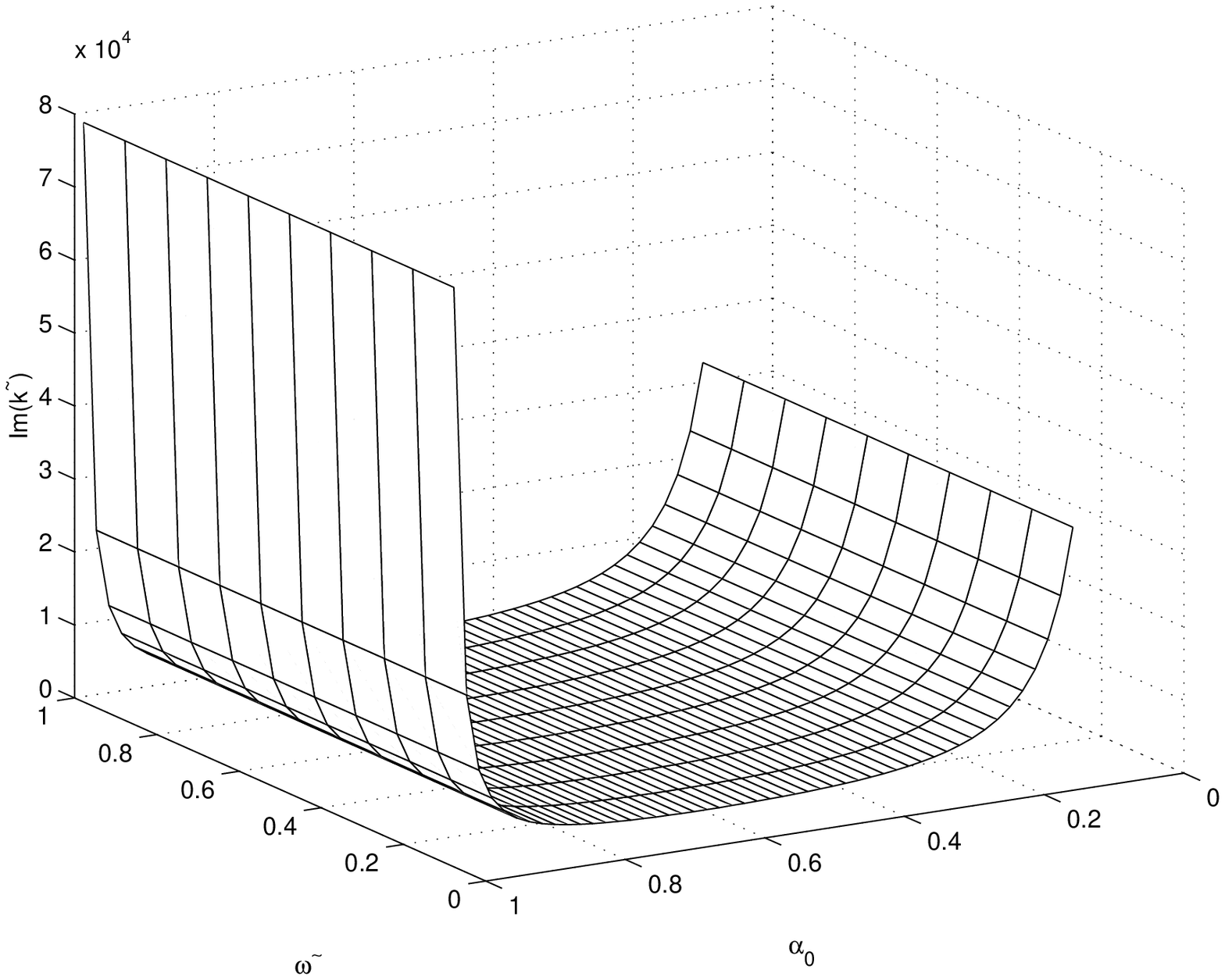}
\end{center}
\caption{\it Top: Real part of the Alfv\'en growth and damping modes for the electron-positron plasma. Center: Imaginary part of the growth mode. Bottom: Imaginary part of the damped mode.}
\end{figure}

For the electron-ion plasma, the ions are essentially nonrelativistic, and the limiting horizon values are chosen to be
\begin{equation}
n_{h1}=10^{18}{\textrm{cm}}^{-3},\quad T_{h1}=10^{10}\textrm{K},\quad n_{h2}=10^{15}{\textrm{cm}}^{-3},\quad T_{h2}=10^{12}\textrm{K}\label{eq71}.
\end{equation}
The limiting horizon value for the equilibrium magnetic field has the same value as it has for the electron-positron case. The limiting horizon temperature for each species has been taken as derived by Colpi et al. \cite{fifty nine} from studies of two temperature models of spherical accretion onto black holes. For the limiting horizon densities and the limiting horizon field, simply arbitrarily those values have been chosen which are consistent with the current ideas. The gas constant and the mass of the black hole have been taken as
\begin{equation}
\gamma _g=\frac{4}{3},\quad M=5M_\odot \label{eq72}.
\end{equation}

\begin{figure}[h]\label{fig2}
\begin{center}
\includegraphics[scale=0.33]{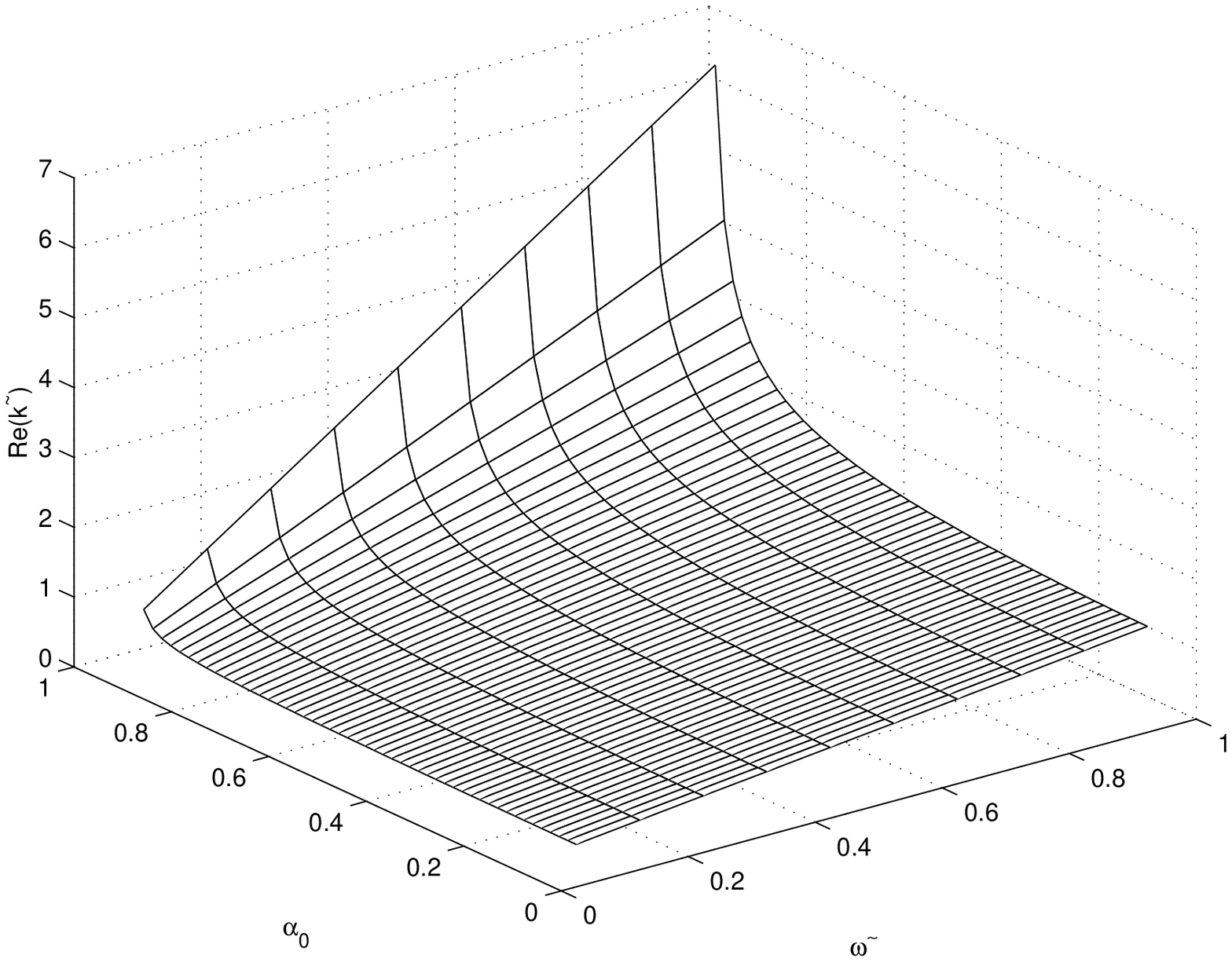} \includegraphics[scale=0.33]{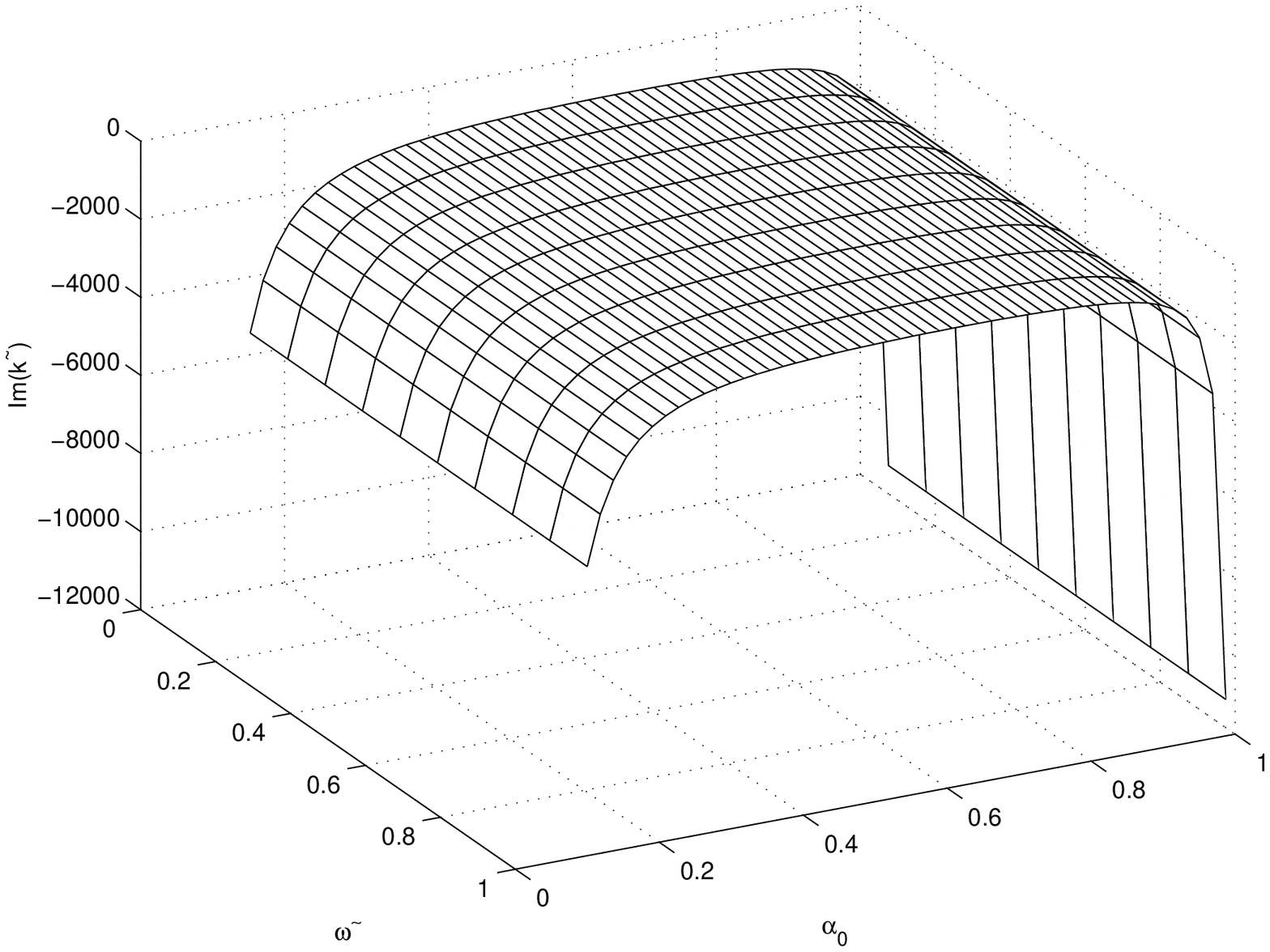}
\end{center}
\caption{\it Left: Real part of the Alfv\'en  growth mode for the electron-ion plasma. Right: Imaginary part of the growth mode.}
\end{figure}

\begin{figure}[h]\label{fig3}
\begin{center}
\includegraphics[scale=0.33]{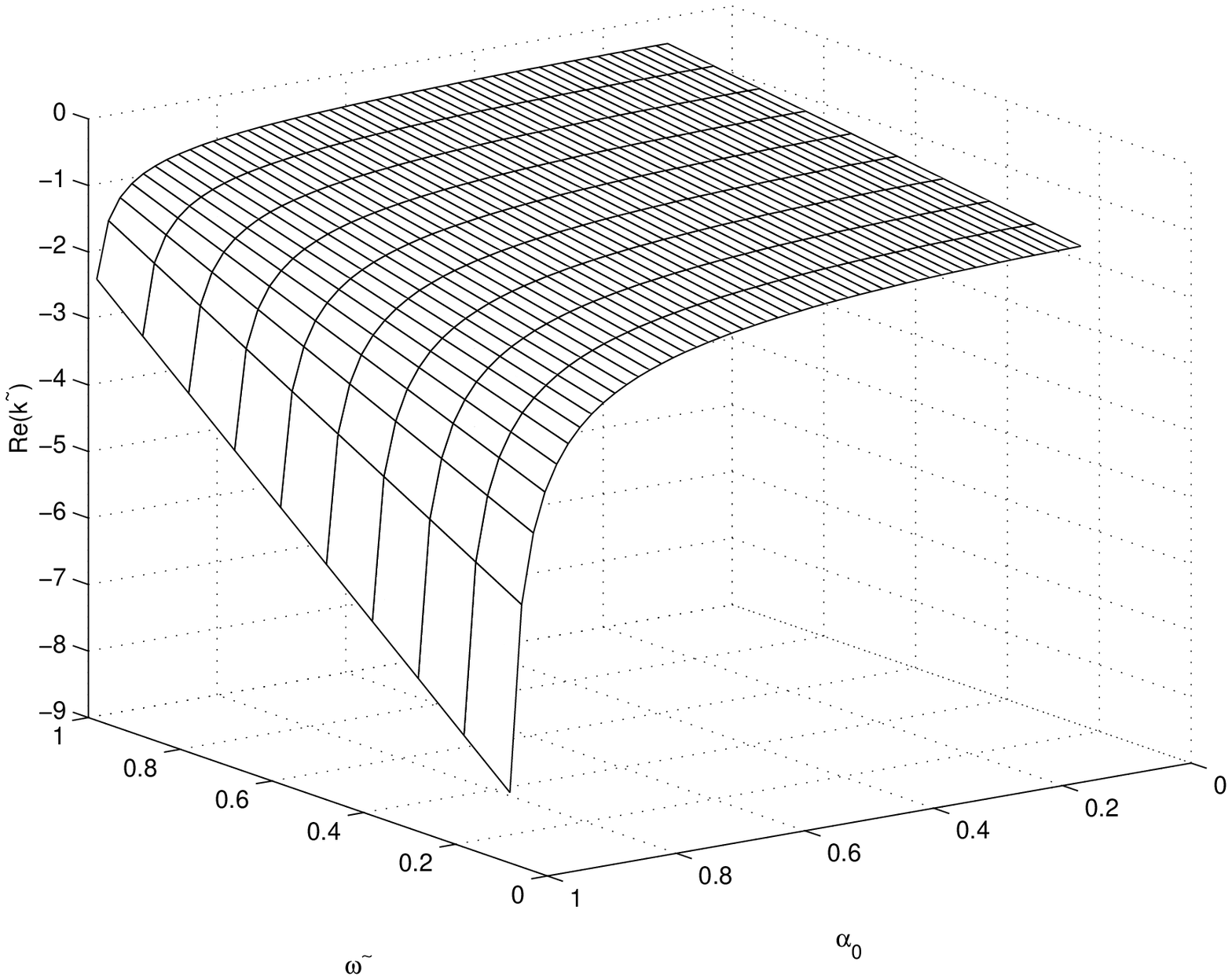} \includegraphics[scale=0.33]{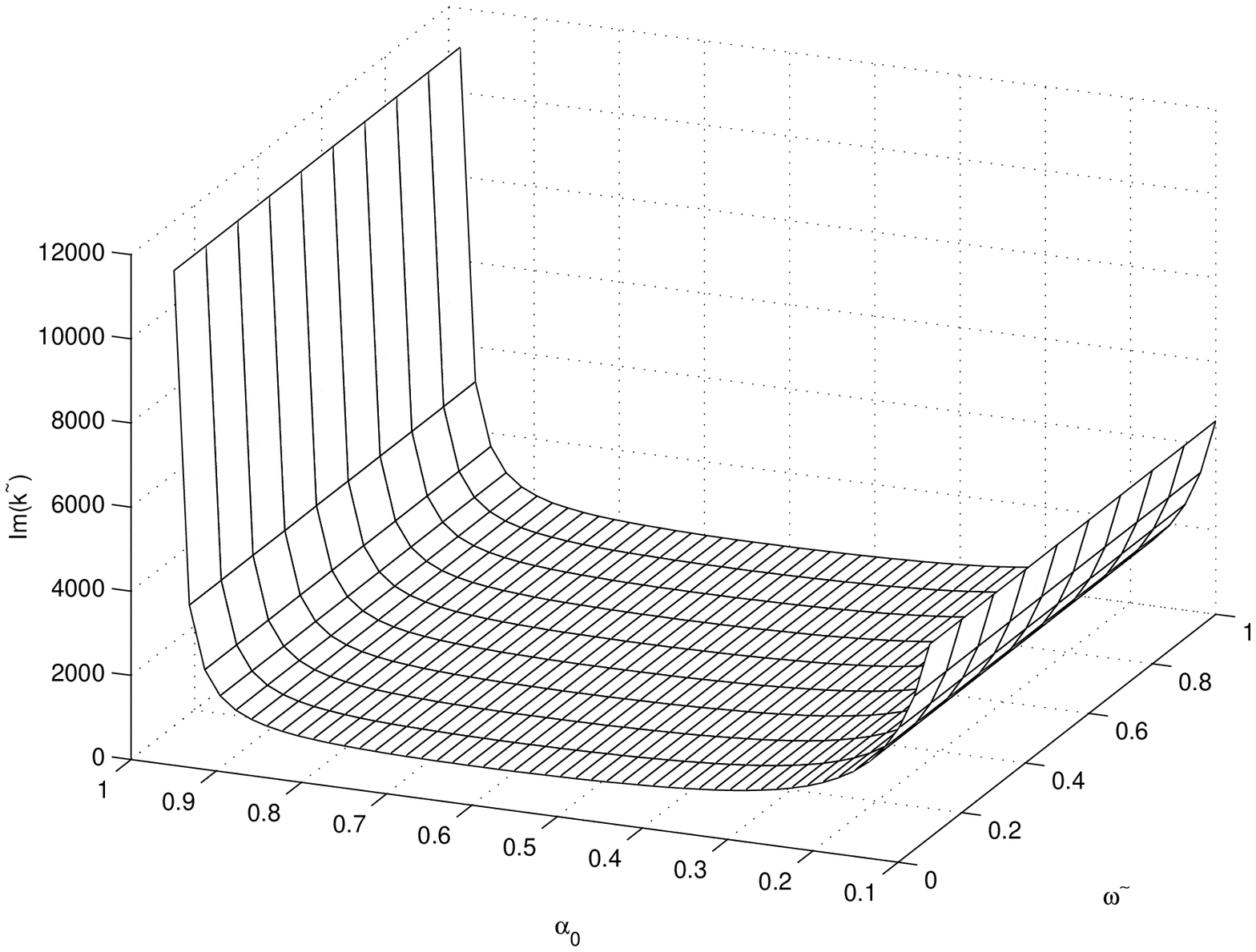}
\end{center}
\caption{\it Left: Real part of Alfv\'en  damped mode for the electron-ion plasma. Right: Imaginary part of the damped mode.}
\end{figure}

\subsection{Alfv\'en Modes}\label{subsec8.1}

\subsubsection{Electron-Positron Plasma}\label{subsubsec8.1.1}
For the ultrarelativistic electron-positron plasma in the special relativistic case, only one real Alfv\'en mode is found to exist \cite{sixty two}, while for the Schwarzschild case there are two Alfv\'en modes \cite{seven}. Our study in the SdS spacetime shows four modes to exist for electron-positron plasma. The first two modes, shown in the Fig. 1 are drawn from the dispersion relation of (\ref{eq67}), (\ref{eq68}) and (\ref{eq69}) and are complex conjugate of each other. They are equivalent to that of Buzzi et al. \cite{seven} in this case. The third mode is damped mode similar to the damped mode of Fig. 1 with larger damping rate and the fourth mode is equivalent to the third mode with opposite in sense and shows growth. These four modes coalesce into a single mode in the special relativistic limit, giving the result of ref. \cite{sixty two}. Since we are using the convention $e^{\textrm{i}kz}=e^{\textrm{i}[\textrm{Re}(k)+ \textrm{iIm}(k)]z}$, the damping corresponds to $\textrm{Im}(\tilde{k})>0 $ and growth to $\textrm{Im}(\tilde{k})<0$.

\subsubsection{Electron-ion Plasma}\label{subsubsec8.1.2}
In this case there exist four modes, two of which are respectively growth and damped as shown in Figs. 2 and 3. The other two modes, complex conjugate to each other, are equivalent to the electron-positron modes shown in Fig. 1 but have larger amount of damping and growth rates. For the first two modes the differences in the magnitudes of the $\omega _{c1}$ and $\omega _{c2}$ apparently lead to take the frequencies from their negative (and therefore unphysical) values for the electron-positron case to positive physical values for the electron-ion case. These changes are thus because of the difference in mass and density factors as between the positrons and ions. It is evident that the growth and damping rates are independent of the frequency, but depended only on the distance from the black hole horizon through $\alpha _o$.

\begin{figure}[h]\label{fig4}
\begin{center}
\includegraphics[scale=0.33]{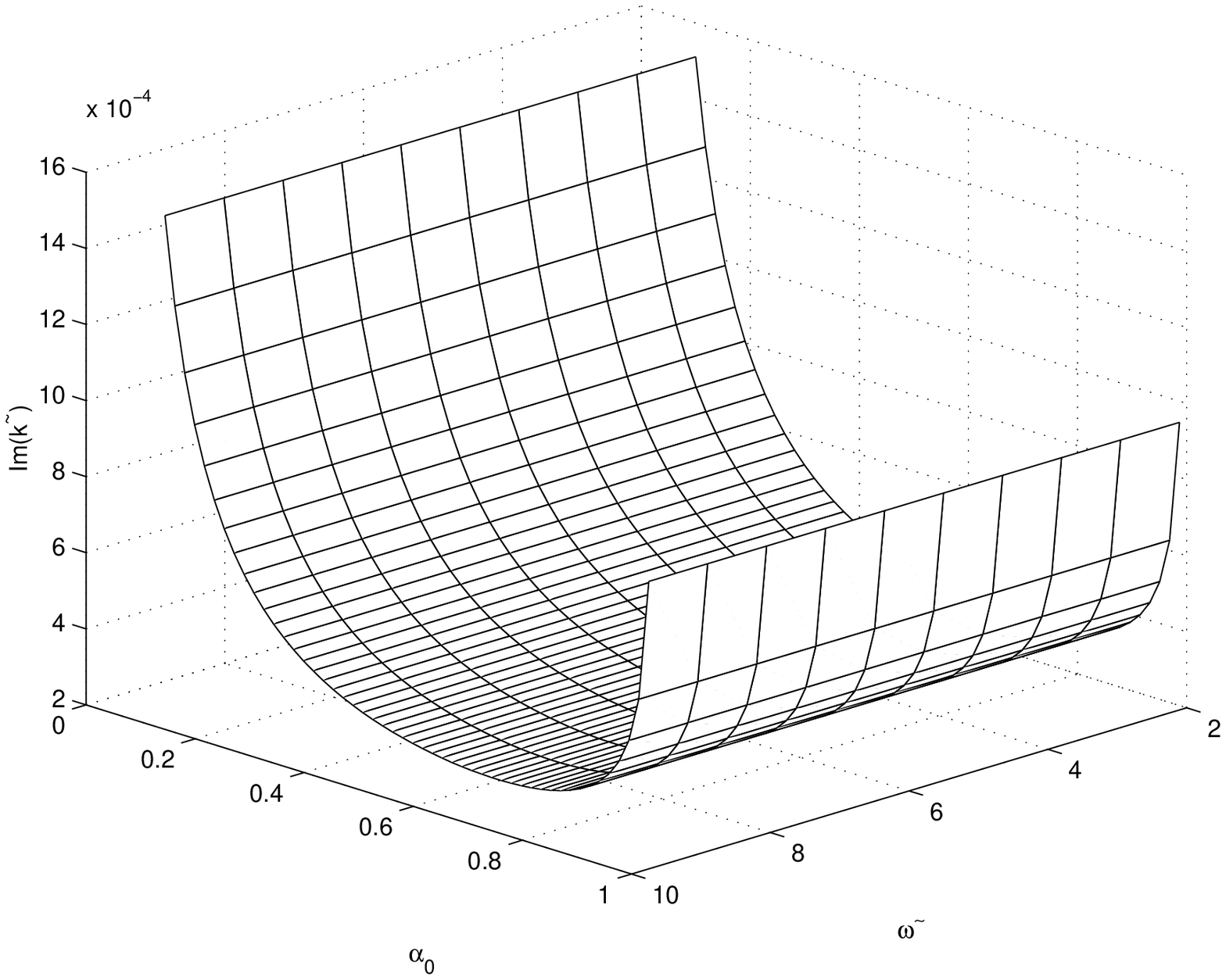} \includegraphics[scale=0.33]{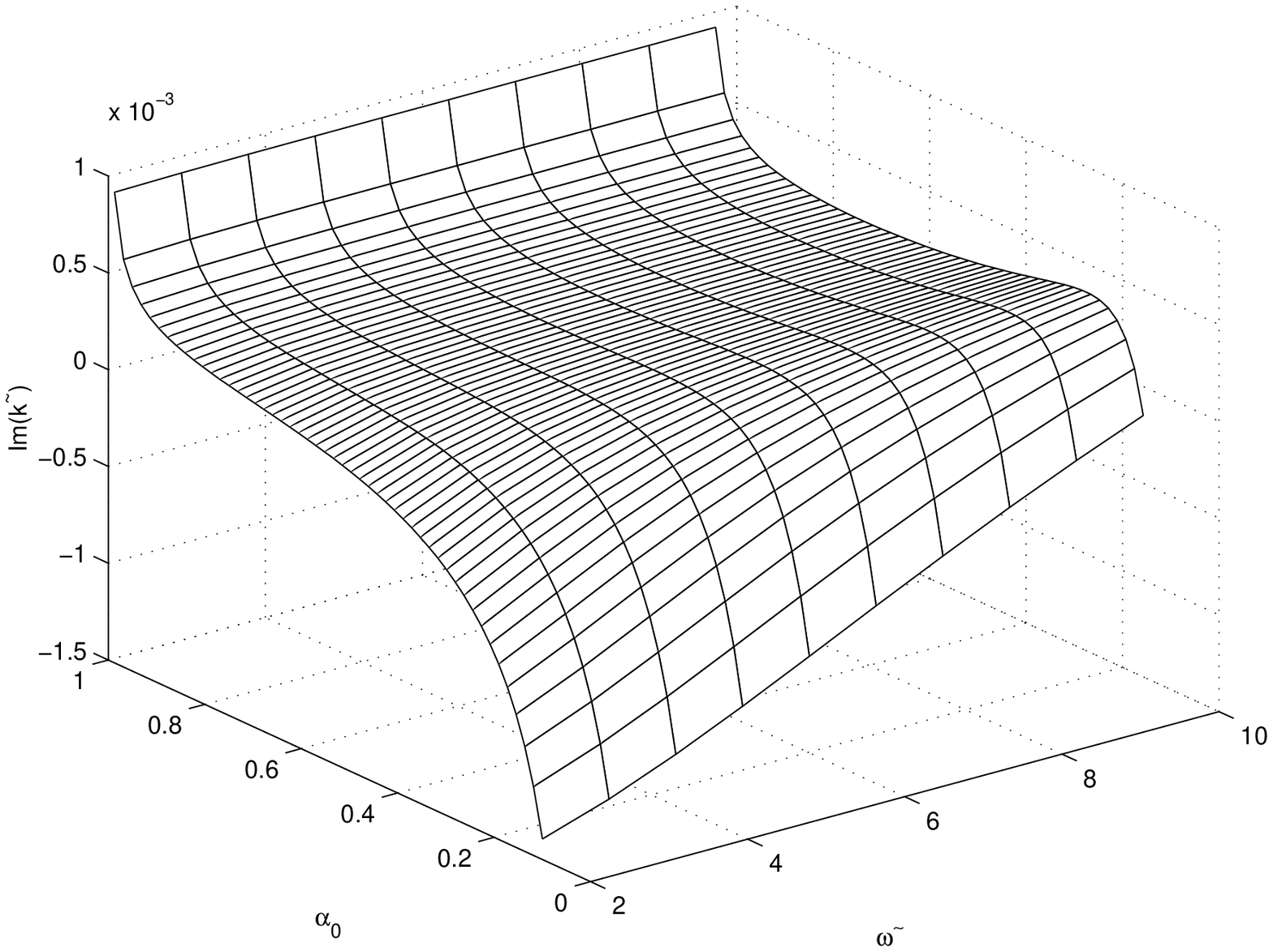}
\end{center}
\caption{\it Left: Imaginary part of high frequency damped mode. Right: Imaginary part of high frequency mode showing both damping and growth.}
\end{figure}

\subsection{High Frequency Transverse Modes}\label{subsec8.2}
\subsubsection{Electron-Positron Plasma}\label{subsubsec8.2.1}
In this case there are four high frequency electromagnetic modes. Two of these modes, shown in Fig. 4, have real parts equivalent to the real part of Fig.1. The left (imaginary) mode in Fig. 4 is damped, while the right one is damped for higher frequency and $\alpha_o>0.2$ and growth for $\alpha_o<0.2$ and lower frequencies. It then appears, at a distance from the horizon corresponding to $\alpha _o<0.2$, that energy is no longer fed into wave mode by the gravitational field but begins to be drained from the waves. The third mode, shown in Fig. 5, is damped for most of the frequency domain and $\alpha_o>0.2$, but shows growth for lower frequencies and $\alpha _o<0.2$. Near to the event horizon, below about $\alpha _o\sim 0.2$, it becomes a growth mode for all frequencies. The fourth mode, shown in Fig. 6, is similar to third mode but shows a growth for all the frequency domain. These four modes reduce to the three modes of Buzzi et al. \cite{seven} in the Schwarzschild case, while they coalesce into only one purely high frequency mode in the special relativistic case \cite{sixty two} for the ultrarelativistic electron-positron plasma.

\begin{figure}[h]\label{fig5}
\begin{center}
\includegraphics[scale=0.33]{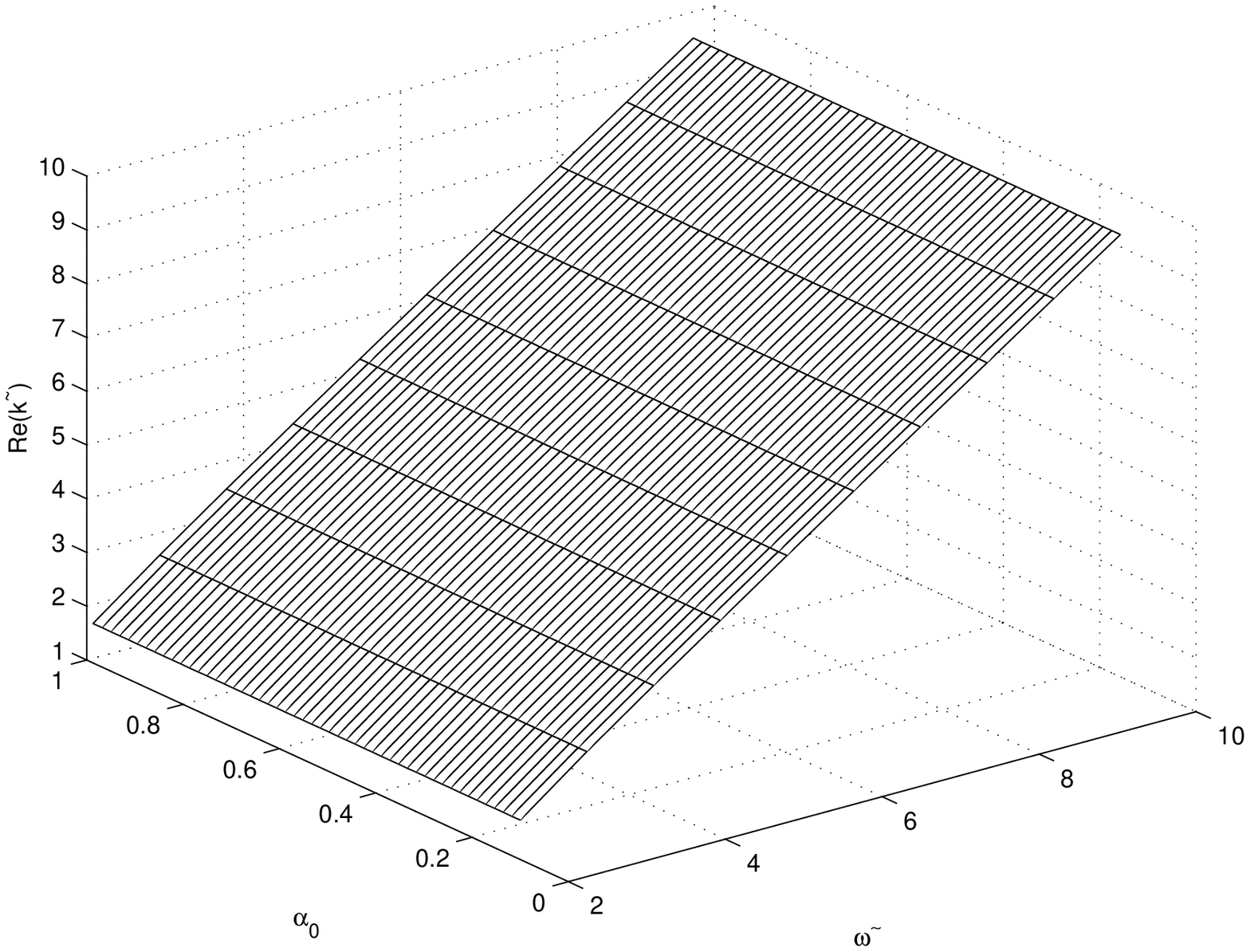}
\includegraphics[scale=0.33]{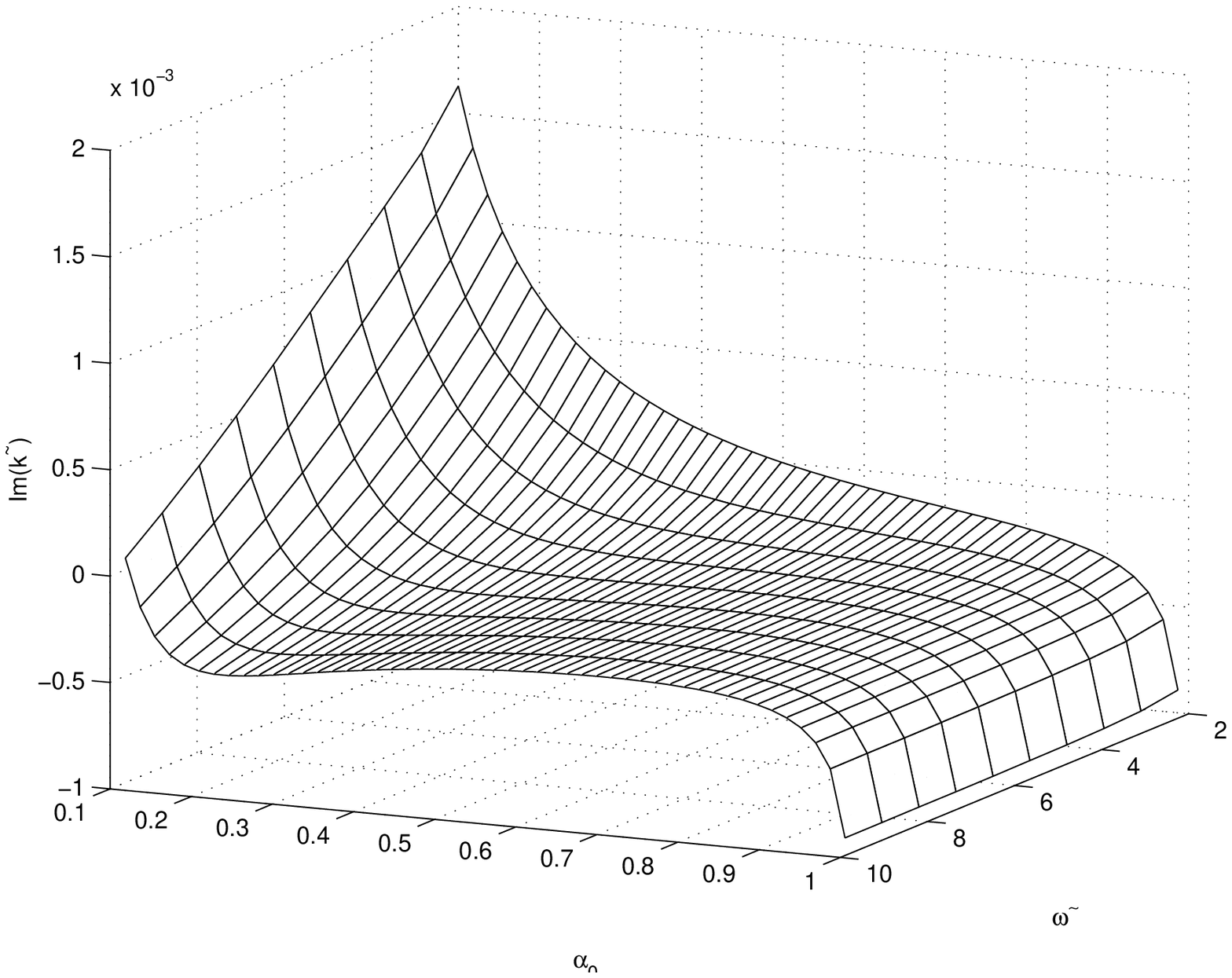}
\end{center}
\caption{\it Left: Real part of high frequency mode for the electron-positron plasma. Right: Imaginary part of high frequency damping and growth mode.}
\end{figure}

\begin{figure}[h]\label{fig6}
\begin{center}
\includegraphics[scale=0.33]{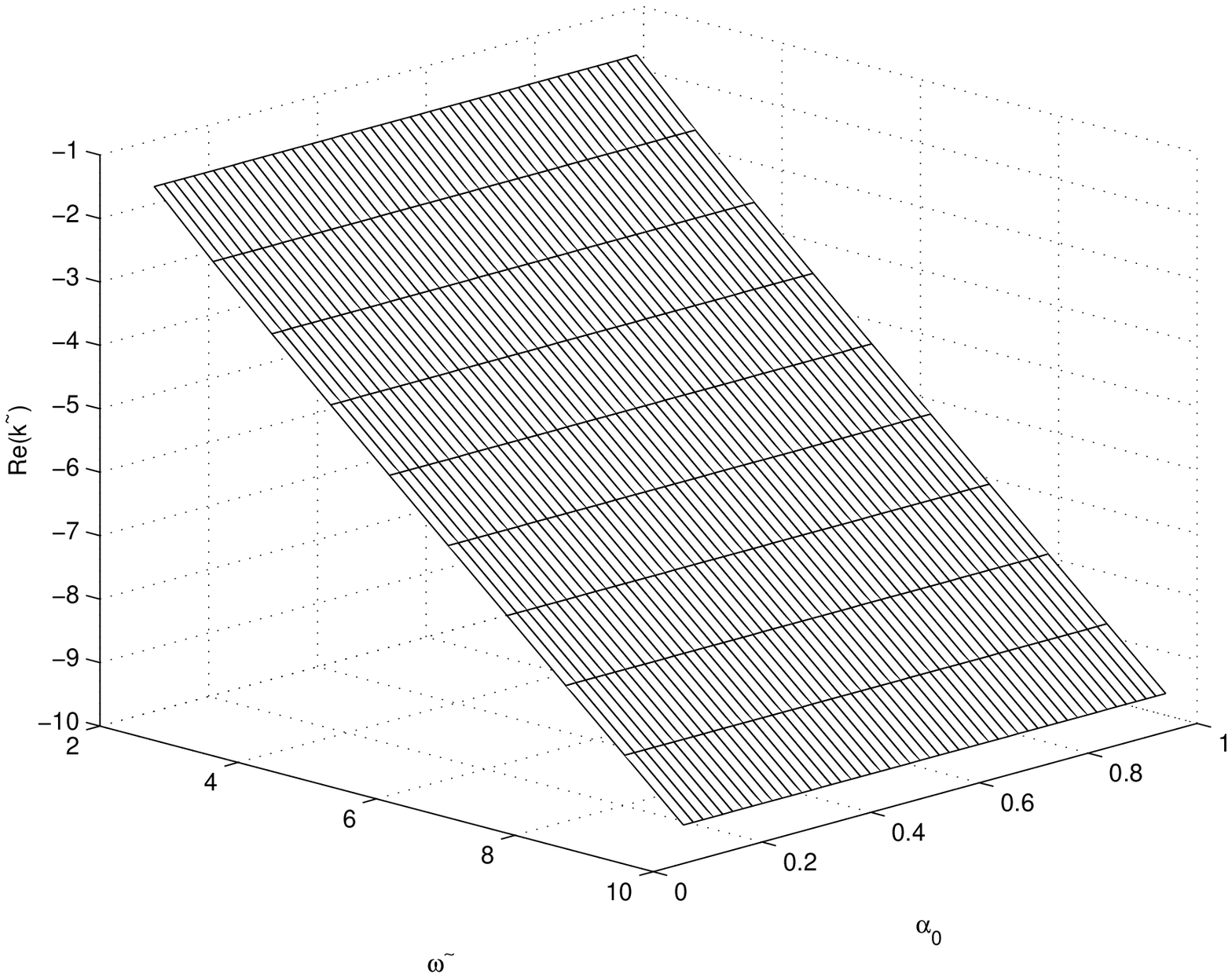}
\includegraphics[scale=0.33]{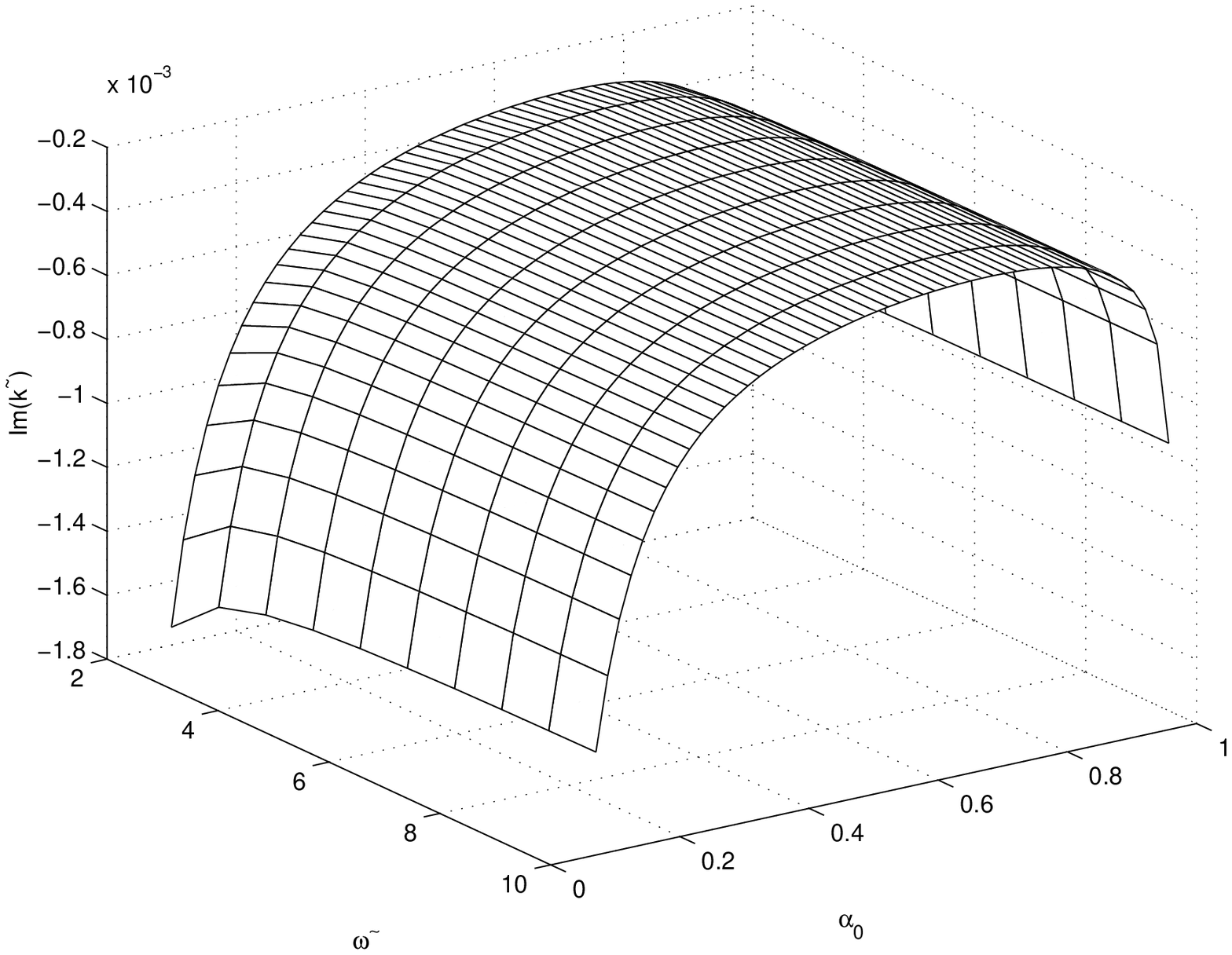}
\end{center}
\caption{\it Left: Real part of high frequency mode for the electron-positron plasma. Right: Imaginary part of high frequency growth mode.}
\end{figure}

\subsubsection{Electron-Ion Plasma}\label{subsubsec8.2.2}
Like the electron-positron plasma, the electron-ion plasma admits four high frequency modes. The first two modes are similar to the modes for electron-positron case shown in Fig. 4, with a large amount of growth and damping rates. The third and fourth modes are similar to the modes shown in Fig. 5 and Fig. 6, respectively, with larger amount of damping and growth. The third mode is growth for higher frequency and $\alpha_o>0.1$, while it is damped for lower frequency and $\alpha_o\rightarrow 0$. The fourth mode is a growth mode and is stable for all frequencies and at all distances from the event horizon. Unlike the corresponding Alfv\'en modes, the growth and decay rates obviously depend on frequency. The four modes reduce to three modes in the Schwarzschild case \cite{seven}.

\section{Concluding Remarks}\label{sec9}
The study of this paper concerns exclusively the investigation, within the local approximation, of Alfv\'en and high frequency transverse electromagnetic waves in a two-plasma surrounding the Schwarzschild black hole in the de Sitter space. Using a local approximation, we have derived the dispersion relations for the Alfv\'en and high frequency electromagnetic waves, and solved it numerically as an analytical solution is unprofitable and not illuminating. In the limit of zero gravity our results reduce to the special relativistic results \cite{sixty two} where only one purely real mode was found to exist for both the Alfv\'en and high frequency electromagnetic waves. In contrast to the special relativistic case, new modes (damped or growth) arise in the black hole spacetime. This is because of the black hole's gravitational field.

The damping and growth rates for the electron-positron plasma are smaller in general, by several orders of magnitude, compared with the real components of the wave number. However, there exist modes for the electron-ion plasma for which the damping and growth rates are significant for the Alfv\'en waves, in particular. For the Alfv\'en waves, the damping and growth rates are obviously frequency independent, but are dependent on the radial distance from the horizon as denoted by the mean value of the lapse function $\alpha _o$. This is of course not the case for the high frequency waves for which the rate of damping or growth is dependent on both frequency and radial distance from the horizon.

Presence of damped modes demonstrates, at least in this approximation, that energy is being drained from some of the waves by the gravitational field. Since the majority of the modes are growth rates, it indicates that the gravitational field is, in fact, feeding energy into the waves.

In the limit $\ell^2\rightarrow\infty$ our study provides the results for the Schwarzschild black hole case \cite{seven}, while for $M=0$ the results reduce to those of the pure de Sitter space \cite{sixty three}. When $M^2=27\ell^2$, the results go for the well known Nariai spacetime. If $\ell^2$ is replaced by $-\ell^2$, our study gives results for the interesting AdS black holes which have received interest in the AdS/CFT correspondence \cite{fifty three,fifty four,fifty five,fifty six} as well as in the context of brane-world scenarios \cite{fifty seven,fifty eight}. In fact, development in string/M theory have greatly stimulated the study of black holes in AdS space. Moreover, recent astrophysical observations indicate that our universe is in a phase of accelerating expansion associated with which is a positive cosmological constant, and the universe therefore might approach a de Sitter phase in the far future \cite{twenty one,twenty four,twenty five,sixty four,sixty five,sixty six,sixty seven}. In view of these reasons, aspects of black holes in the de Sitter space might be of interest, and our study of two-fluid plasma in this paper near the event horizon of the Schwarzschild-de Sitter black hole is thus well motivated.

\vspace{1.0cm}

\noindent
{\large\bf Acknowledgement}\\
One of the authors (MHA) thanks the SIDA as well as the Abdus Salam International Centre for Theoretical Physics (ICTP), Trieste, Italy, for supporting with an Associate position of the Centre.

\end{document}